\documentclass[superscriptaddress,prc,twocolumn,showpacs,preprintnumbers,amsmath,amssymb,altaffilletter]{revtex4}
\usepackage{graphicx}
\usepackage{xspace}
\usepackage{sidecap}
\usepackage{amsmath}
\usepackage{amssymb}
\usepackage{url}
\usepackage{braket} 
\usepackage{slantsc}
\usepackage[symbol]{footmisc}
\usepackage[T1]{fontenc}
\usepackage[usenames]{color}

\def\nuc#1#2{${}^{#1}$#2}

\def\BBz{$\beta\beta(0\nu)$}

\def\BBt{$\beta\beta(2\nu)$}

\def\BB{$\beta\beta$}

\def\gA{$g_{A}$}                  
\def\be{\begin{equation}}
\def\ee{\end{equation}}

\begin{document}
\title{Solar neutrino interactions with the double beta decay nuclei of $^{82}$Se, $^{100}$Mo and $^{150}$Nd 
}


\newcommand{\lanl}{Los Alamos National Laboratory, Los Alamos, NM, USA}
\newcommand{\ou}{Research Center for Nuclear Physics, Osaka University, Ibaraki, Osaka 567-0047, Japan}

\affiliation{\ou} 
\affiliation{\lanl} 

\author{H.~Ejiri}\affiliation{\ou}
\author{S.R.~Elliott}\affiliation{\lanl}

\begin{abstract}
Solar neutrinos interact within double-beta decay (\BB) detectors and contribute to backgrounds for \BB\ experiments. Background contributions due to solar neutrino interactions with \BB\ nuclei of  $^{82}$Se, $^{100}$Mo, and $^{150}$Nd are evaluated. They are shown to be significant for future high-sensitivity \BB\ experiments that may search for Majorana neutrino masses in the inverted-hierarchy mass region. The impact of solar neutrino backgrounds and their reduction are discussed for future \BB\ experiments.   \\

\end{abstract}
\maketitle

\section{Introduction}
Neutrino-less double beta decay (\BBz) is a unique and realistic probe for studies of neutrino ($\nu$) properties and especially the Majorana mass character of the neutrino and the absolute mass scale. \BB\ studies and $\nu$ masses are discussed in recent reviews and their references \cite{ell04,eji05,avi08,ver12}. 

The rate of \BBz, if it exists, would be extremely small because \BB\ is a second-order weak process that requires lepton number conservation violation and Majorana fermions. There are numerous possible mechanisms that can give rise to \BBz\ as is discussed in the reviews\cite{ell04,eji05,avi08,ver12}. It is known, however, that light neutrinos exist and therefore it is convenient to consider light-neutrino exchange as the default process by which to benchmark \BBz\ rates. For light-neutrino exchange, the rate depends on the effective Majorana mass squared and the typical mass regions to be explored are about 45-15 meV and 4-1.5  meV in cases of the inverted-hierarchy (IH) and normal-hierarchy (NH) mass regions, respectively. The \BBz\ half-lives expected for these regions are near or greater than $10^{27}$ years, depending significantly on the nuclear matrix element (NME), including the effective axial weak coupling \gA. As a result, the \BBz\  signal rate ($S_{\beta \beta}$) is near or less than a few counts per ton of \BB\ isotope per year (t y).  Accordingly the background rate necessarily has to be around or less than one count per t y.  

Solar-$\nu$s are omnipresent and cannot be shielded, and thus their charged current (CC) and neutral current (NC) interactions are potential background sources for high sensitivity \BB\ experiments as discussed in \cite{eji14,bar11} and references therein. In fact, it has been shown that solar-$\nu $ CC interactions with \BB\ isotopes like $^{100}$Mo~\cite{eji00}, $^{116}$Cd~\cite{zub03} and $^{150}$Nd~\cite{zub12} can be used for real-time studies of the low-energy solar-$\nu$s. 

The \BB\ isotopes most often used or considered for high-sensitivity experiments are $^{76}$Ge, $^{82}$Se, $^{100}$Mo, $^{130}$Te, $^{136}$Xe and $^{150}$Nd.  Here, we classify them into two groups. Group A consisting of $^{82}$Se, $^{100}$Mo  and $^{150}$Nd have a large solar-$\nu $ CC rate, whereas Group B consisting of $^{76}$Ge, $^{130}$Te and $^{136}$Xe have a rather small solar-$\nu$ CC rate. In the previous paper~\cite{eji14}, background contributions from CC solar-$\nu$ interactions were discussed for the 3 \BB\ nuclei in Group B. Solar-$\nu$ interactions with atomic electrons in \BB\ isotopes and liquid scintillators used for \BB\ experiments were considered in Refs.~\cite{ell04,bar11,bar14,eji16}. 
 
The present paper aims to evaluate the background contributions for the 3 nuclei in Group A and discuss the impact on high sensitivity \BB\ experiments using them.  The mechanics of these calculations are identical to those in Ref.~\cite{eji14}, and we do not repeat all the details here.

\section{Solar Neutrino Backgrounds}

\begin{figure}[htb]
\caption{Schematic diagrams of the \BB\ of $^{Z-1}$A to $^{Z+1}$A, the solar-$\nu $ CC interaction on $^{Z-1}$A and the electron ($\gamma $) 
decay to $^{Z}$A,  and  the single $\beta $/$\gamma $ decays of $^{Z}$A to $^{Z+1}$A. $Q_{\beta \beta }$, $Q_{\beta}$ , $Q_{\nu}$, and $Q_{e}$  are the $Q$
values for ${\beta \beta }$, $\beta $, $\nu $ CC and electron capture (EC), respectively. 
\label{fig:solar}}
\begin{center}
\includegraphics[width=0.5\textwidth]{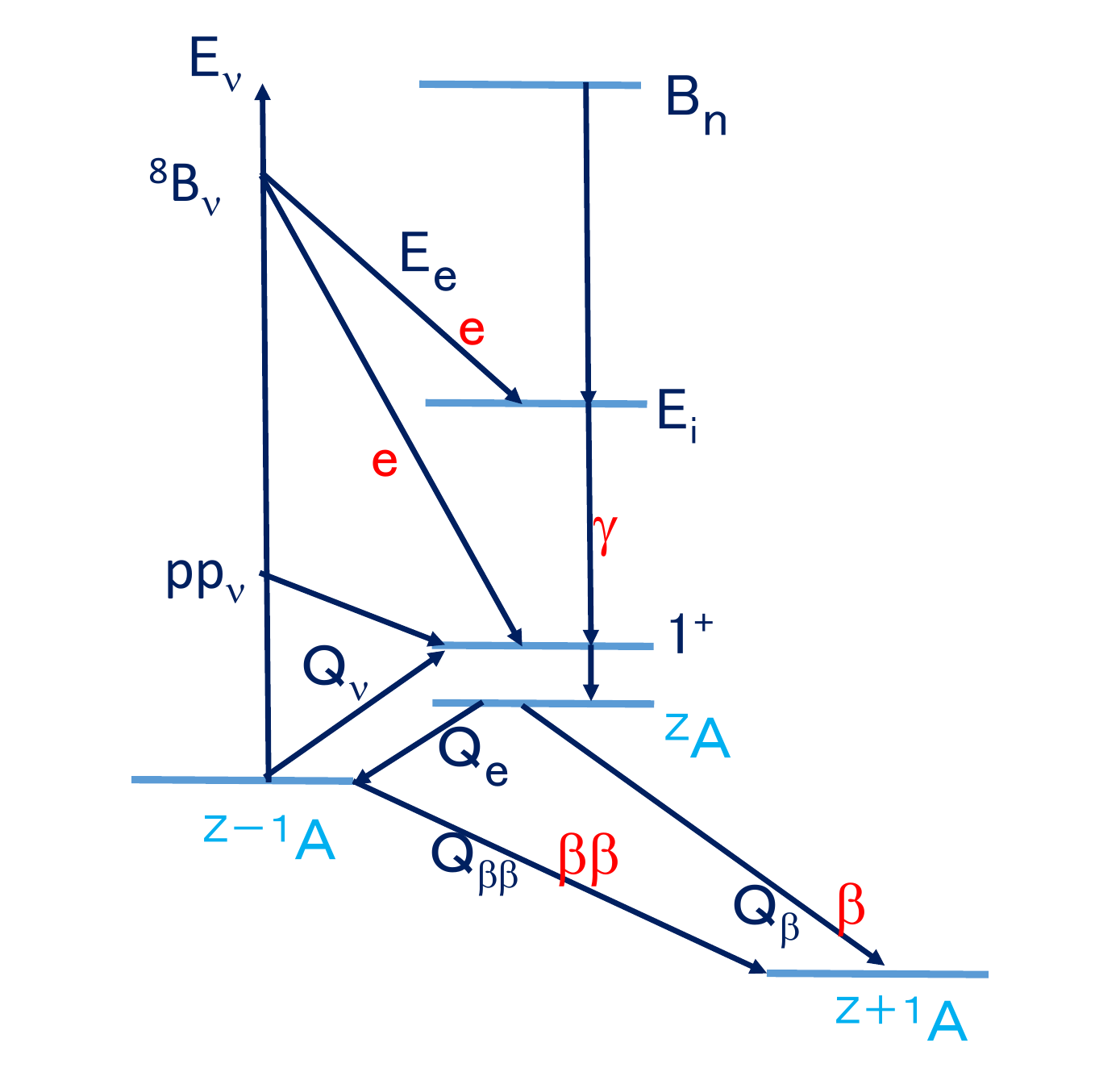}
\end{center}
\end{figure}

The process of \BB\ decay from $^{Z-1}$A to  $^{Z+1}$A via the intermediate nucleus $^{Z}$A is shown in Eqn.~\ref{eqn:Processes}. Solar $\nu$s can produce background to this signal primarily through CC interactions with \BB\ nuclei. The CC interaction produces background in two ways. First the CC interaction itself can produce a signal ($B_{CC}$) given by the promptly emitted e$^-$ and, if the resulting nucleus is in an excited state, a number of $\gamma$ rays may be emitted as the nucleus relaxes to its ground state. Second, the resulting nucleus $^{Z}$A can then $\beta^-$ decay to $^{Z+1}$A by emitting a single $\beta^-$ ray and possibly also $\gamma $ ray(s) if the residual state is an excited state ($B_{SB}$). The interaction and decay schemes are shown in Fig.~\ref{fig:solar}. The 3 processes are expressed as,
\begin{eqnarray}
\beta\beta:~~  ^{Z-1}A 		&   \rightarrow	& ^{Z+1}A+ \beta^- +\beta^- +Q_{\beta \beta } \nonumber	\\
CC:~~		 ^{Z-1}A+\nu	&  \rightarrow 	& ^{Z}A +e^- +\gamma(s) +Q_{\nu}\nonumber	\\
SB:~~		 ^{Z}A		&  \rightarrow 	& ^{Z+1}A  + \beta^- +\gamma(s) +Q_{\beta},
\label{eqn:Processes}
\end{eqnarray}
where \BB, CC, and SB denote the double beta decay, the solar-$\nu$ CC interaction, and the single beta decay processes, respectively. The $Q$ values for each are given as $Q_{\beta \beta }$, $Q_{\nu}$, and $Q_{\beta }$ respectively, as shown in Fig.~\ref{fig:solar}. 

We consider \BB\ detectors where the sum energy of the $\beta $ and $\gamma $ rays is measured. The \BBz\ signature is a peak within the region of interest (ROI) at the \BB\ Q value ($E=Q_{\beta \beta }$).  In contrast, the sum of the electron energy ($E_e$) and any successive $\gamma $ ray energy ($E_{\gamma }$) is a continuum spectrum for both the CC and SB processes. The backgrounds of interest are estimated by their yields within the ROI and therefore are sensitive to the detector's energy resolution.  The energy width of the ROI is given by the FWHM resolution ($\Delta E$) at the energy of the Q value. This ratio, defined as $\delta \def \Delta E/Q_{\beta \beta }$, is the fractional energy resolution at $E=Q_{\beta \beta }$. Certainly the kinematics and event topologies for the backgrounds discussed here may permit experimental techniques for rejection that go beyond just a sum energy cut. We cannot anticipate all possible rejection techniques, but instead estimate rates so that future efforts may better assess these backgrounds for a given experimental configuration. In fact, methods to reduce the solar-$\nu$ backgrounds do depend on the detector configuration and the $\beta^-$ and $\gamma$ decay scheme of the nucleus. We briefly discuss these later for individual nuclei.

\subsection{The \BBz\ Rate}
We first evaluate the \BBz\ signal rate $S_{\beta\beta}$ for the light $\nu$-mass process and then compare to estimated CC and SB background rates. Although there are numerous other potential \BBz\ mechanisms, providing this rate estimate enables a comparison to the background rates. The \BBz\ signal rate for the light Majorana-$\nu $ exchange is written as~\cite{eji05,ver12}
\begin{equation}
S_{\beta\beta} = ln2 ~G_{0\nu}(m_{eff})^2 [M_{0\nu}]^2 \epsilon_{\beta\beta} \frac{6 \times 10^{29}}{A}~ /\mbox{t y},
\end{equation}
where $S_{\beta\beta}$ is the signal rate per ton per year (t y) of the \BB\ isotope,  $G_{0\nu}$ is the phase space volume, $m_{eff}$ is the effective Majorana $\nu $-mass in units of the electron mass, $\epsilon_{\beta\beta}$ is the \BBz\ peak detection efficiency, and $M_{0\nu}$ is the nuclear matrix element (NME) for the light $\nu$-mass process. Here $G_{0\nu}$ includes conventionally the axial weak coupling $g_A=1.267g_V $ with $g_V$ being the vector coupling constant, and $M_{0\nu}$ is given by the sum of the axial-vector and the vector NMEs. 

The NMEs are expressed as $M_A=(g^{eff}_A/ g_A)M_A^m$ and  $M_V=(g^{eff}_V/ g_V)M_V^m$, where $M_A^m$ and $M_V^m$ are model NMEs and  ($g^{eff}/g$) are  the renormalization (quenching) factors due to such non-nucleonic (isobar, exchange current etc) and nuclear medium effects that are not explicitly included in the model NMEs \cite{eji05,ver12,eji78}. In a typical case of $G_{0\nu}= 5 \times 10^{-14}$/y, $m_{eff}$ = 20 meV/$m_e$, $M_{0\nu}$=2, $A$=100, and $\epsilon_{\beta\beta}$=0.6, the \BBz\ signal rate is $S_{\beta\beta}\approx 1$/t~y.
 
 \subsection{The $\beta$ Decay of $^{Z}$A}
Next, we evaluate the background rate for the SB case. For SB, all solar-$\nu$ sources exciting the intermediate states below the neutron threshold energy $B_{n}$ in $^{Z}$A contribute to the production of the ground state of $^Z$A via $\gamma$ decay. Hence the production rate for $^{Z}$A is given by the total solar-$\nu$ capture rate in units of SNU ($S_t$). The background rate per ton year for SB ($B_{SB}$) is expressed as
\begin{equation}
B_{SB}= 3.15\times10^{-29} n_{\beta\beta} S_{t} \epsilon_{SB},
\end{equation}
where  $n_{\beta\beta}$ is the number of \BB\ isotope nuclei per ton, and $\epsilon_{SB}$ is the effective efficiency for the SB signal being located at the ROI after various cuts. 

The solar-$\nu$ capture rates for individual neutrino sources are evaluated by using the neutrino responses ($B(GT), B(F)$) given by recent charge exchange reactions \cite{fre16,thi12,gue11} and the neutrino fluxes from BP05(OP)~\cite{bah05}. The CC neutrino responses, $B(GT)$, have been studied by using high energy-resolution experiments at RCNP Osaka. The calculations were done as in Ref.~\cite{eji14} including the treatment of $\nu$ oscillations. The \BB, solar-$\nu$ interaction, and single $\beta$ $Q$ values and the solar-$\nu$ capture rates for Group A nuclei are shown together with those for Group B nuclei \cite{eji14} in Table~\ref{tab:SolarRates}.
 
\begin{table*}[htpb]
\caption{\BB, CC, and SB $Q$ values in units of MeV and solar-$\nu $ capture rates in units of SNU for selected \BB\ nuclei including the effect of oscillations.  Column 8 gives $S_t$ for no oscillations. $Q_{\beta \beta}$ is the \BB\ $Q$ value, $Q_{\nu}$ is the $\nu$-CC $Q$ value for the lowest 1$^+$ state, $Q_{\beta}$ is the single $\beta$ Q value, $S_{B}$ is the $^8$B-$\nu$ capture rate, and $S_{t}$ is the total solar-$\nu$ capture rate. The background rates for $\beta$ decay ($B_{SB}$) and \BBt\ ($B_{2\nu}$) are calculated for $\delta = 0.02$. The small differences in solar-$\nu$ capture rates in this table compared those reported in Ref.~\cite{eji14} are due to a small arithmetic error in the \nuc{7}{Be} flux calculations in that previous paper.
\label{tab:SolarRates}}
\centering
\vspace{0.5cm}
\begin{tabular}{ccccccccccc}
\hline
Isotope 	& \BBt\ $\tau_{1/2}$				& $Q_{\beta \beta }$ 	& $Q_{\nu}$ 	& $Q_{\beta}$  & $S_{pp}$ 	& $S_{B}$  & $S_t$ no osc. 	&$S_t$   		& $B_{SB}$ 	& $B_{2\nu}$ \\
		& years 						& MeV 			& MeV 		&  MeV 		&  (SNU)  		& (SNU) 	& (SNU) 			& (SNU) 		& events/ t y 	& events /t y \\
\hline\hline
$^{82}$Se & $9.2\times10^{19}$~\cite{bar15}	& 2.992 			&~~ -0.172		& ~~3.093		& 257 		& 10.0	& 672  			&  368		& 4.42		&0.15\phantom{0} \\
$^{100}$Mo & $7.1\times10^{18}$~\cite{bar15}	& 3.034 		&~~ -0.168 		& ~~3.202 		& 391 		&\phantom{0}6.0 & 975 		&  539		&  0.11		&1.56\phantom{0}\\
$^{150}$Nd & $8.2\times10^{18}$~\cite{bar15}	& 3.368 		&~~ -0.197 		& ~~3.454 		& 352 		&15.5	&  961  			&  524		& 0.12		&1.00\phantom{0}  \\
$^{76}$Ge & $1.93\times10^{21}$~\cite{ago15} 	& 2.039 		&~~ -1.010 		& ~~2.962 		& 0 			&  \phantom{0}5.0  & \phantom{0}15.7  	& \phantom{00}6.3 & 0.03		&0.005 \\
$^{130}$Te & $6.9\times10^{20}$~\cite{bar15}	& 2.528 		& ~~-0.463		& ~~2.949 		& 0			&\phantom{0}6.1   &  \phantom{0}67.7 		&  \phantom{0}33.7 		&0.48		&0.01\phantom{0} \\
$^{136}$Xe &$2.19\times10^{21}$~\cite{bar15}	& 2.468 		& ~~-0.671 		& ~~2.548 		& 0			&\phantom{0}9.8& 136 		&   \phantom{0}68.8		& 0.55		&0.003\\
\hline\hline
\end{tabular}
\end{table*}

\begin{figure}[htb]
\caption{Solar-$\nu $ capture rates in units of SNU for current \BB\ nuclei  are plotted against the neutrino CC $Q_{\nu}$ value for the lowest 1$^+$ state.  $S_{t}$ is the total capture rate. Group A nuclei have a large $S_{t}$. Group B nuclei have a small  $S_{t}$.
\label{fig:rates}. }
\begin{center}
\includegraphics[width=9 cm]{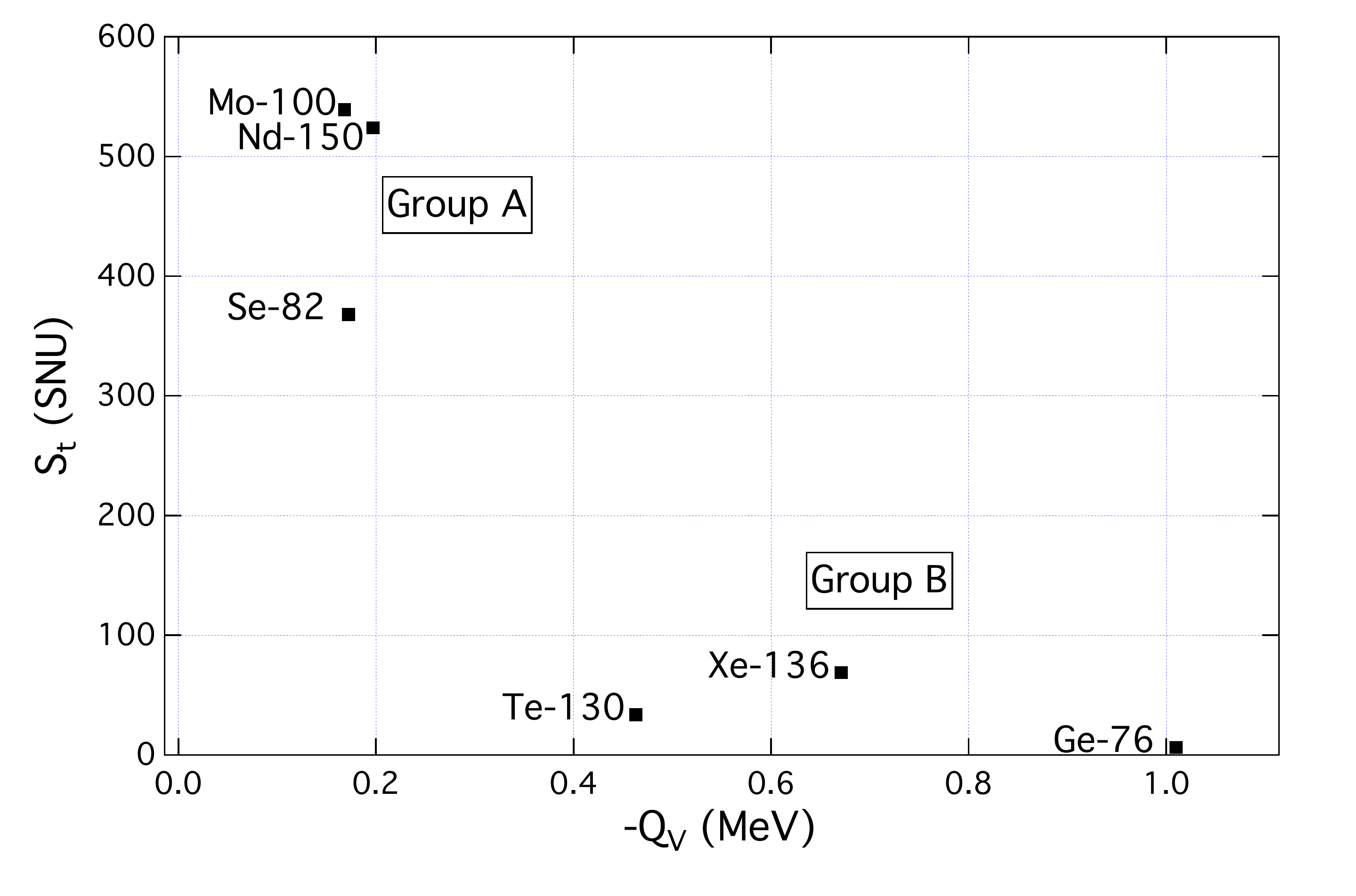}
\end{center}
\end{figure}

$S_t$ for Group A and B nuclei are plotted against the neutrino CC $Q_{\nu}$ value in Fig.~\ref{fig:rates}. The Group A nuclei of $^{82}$Se, $^{100}$Mo  and $^{150}$Nd with small negative $Q_{\nu}$ values around -170 keV, have large solar-$\nu$ capture rates because they are strongly excited by the pp neutrinos. On the other hand Group B nuclei of $^{76}$Ge, $^{130}$Te and $^{136}$Xe have rather small solar-$\nu$ CC rates because the threshold energy (-$Q_{\nu}$) is large enough that the pp neutrinos can not excite them. The pp neutrino contributions to the total capture rates for the Group A nuclei  are around 60\% of the total, while the $^7$Be and $^8$B neutrino capture rate are around 30\% and 3\%, respectively. 
 
$\epsilon _{SB}$ is evaluated as a function of $\delta$ for simple calorimetric detectors. The background rates are approximately proportional to the resolution, i.e. the width of the ROI. $B_{SB}$ for a fractional resolution of $\delta =0.02$ are given in Table~\ref{tab:SolarRates}. 

For the cases of $^{100}$Mo and $^{150}$Nd, the ROI is located near the end-point energy at the tail of the single $\beta$-ray spectrum because $Q_{\beta\beta}$ is very close to $Q_{\beta}$. As a result, $\epsilon_{SB}$ is much reduced and $B_{SB}$ is small for these nuclei.  In the case of $^{82}$Se, however, the intermediate nucleus $^{82}$Br decays primarily to the highly excited state at 2.648 MeV. Thus the ROI is located at the middle of the single $\beta$-ray energy spectrum resulting in $\epsilon_{SB}$ being relatively large, and a correspondingly large $B_{SB}$.  $B_{SB}$ for the Group A  nuclei are also given in Table~\ref{tab:SolarRates}.

\subsubsection{\nuc{82}{Se}}
\begin{figure}[htb]
\caption{The sum spectrum of $\beta$ and $\gamma$-ray energies for the decay of \nuc{82}{Br}. The hatched region shows the ROI fraction for a $\delta = 0.02$.
\label{fig:Br82}. }
\begin{center}
\includegraphics[width=9 cm]{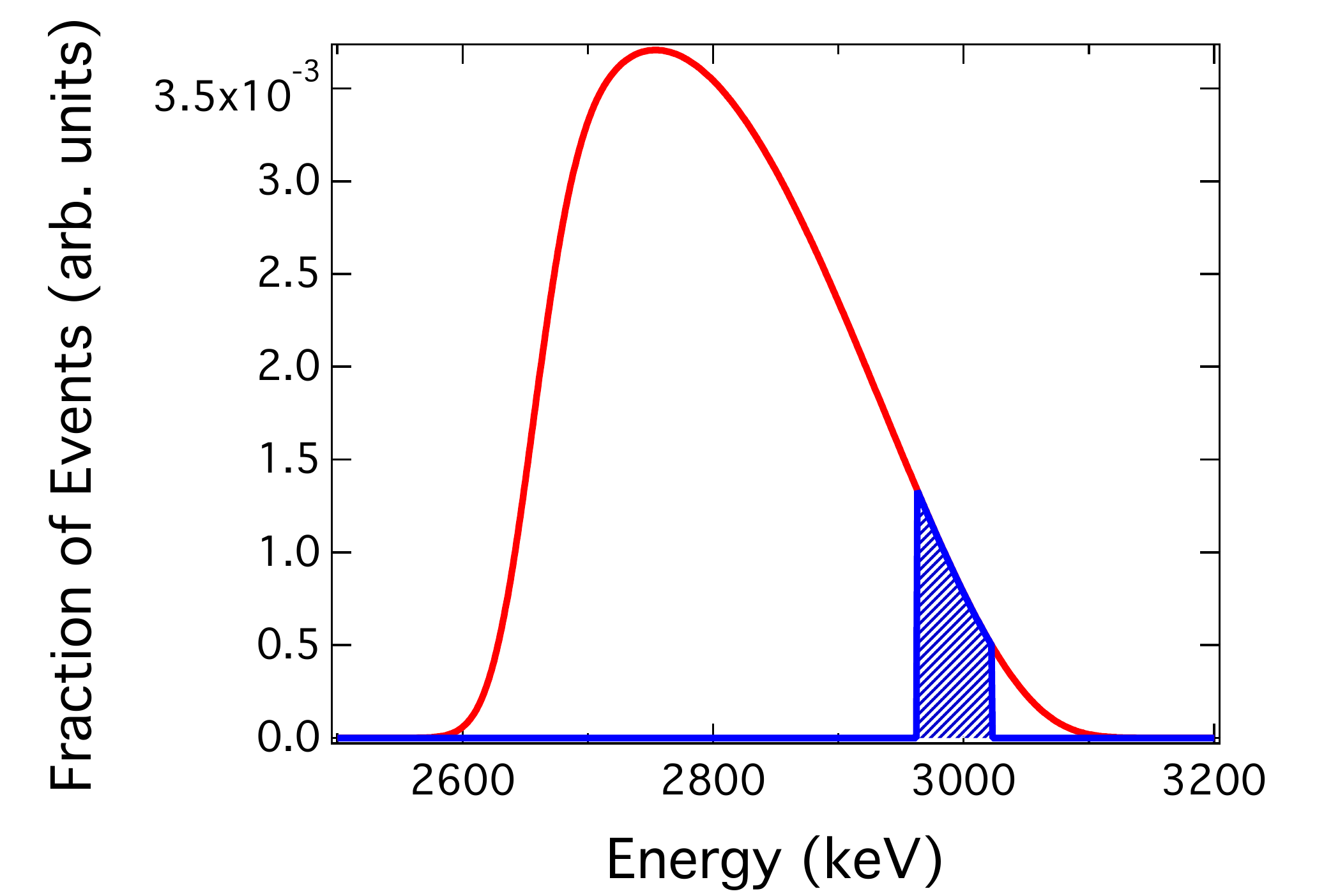}
\end{center}
\end{figure}

The oscillated solar $\nu$ capture rate on \nuc{82}{Se} is calculated to be 368 SNU.  \nuc{82}{Br} decays with a 98.5\% branching ratio to the 2.648-MeV state in \nuc{82}{Kr} and this was the only state considered in our calculations. The $\epsilon_{SB}$ for decay of the \nuc{82}{Br} product to populate the ROI was calculated to be 2.6\%, 5.2\% and 15.7\% for resolutions of $\delta$ = 0.01, 0.02 and 0.05 respectively. The resulting $SB$ spectrum is shown in Fig.~\ref{fig:Br82}.

\subsubsection{\nuc{100}{Mo}}
\begin{figure}[htb]
\caption{The sum spectrum of $\beta$ and $\gamma$-ray energies for the decay of \nuc{100}{Tc}. The hatched region shows the ROI fraction for a $\delta = 0.02$.
\label{fig:Tc100}. }
\begin{center}
\includegraphics[width=9 cm]{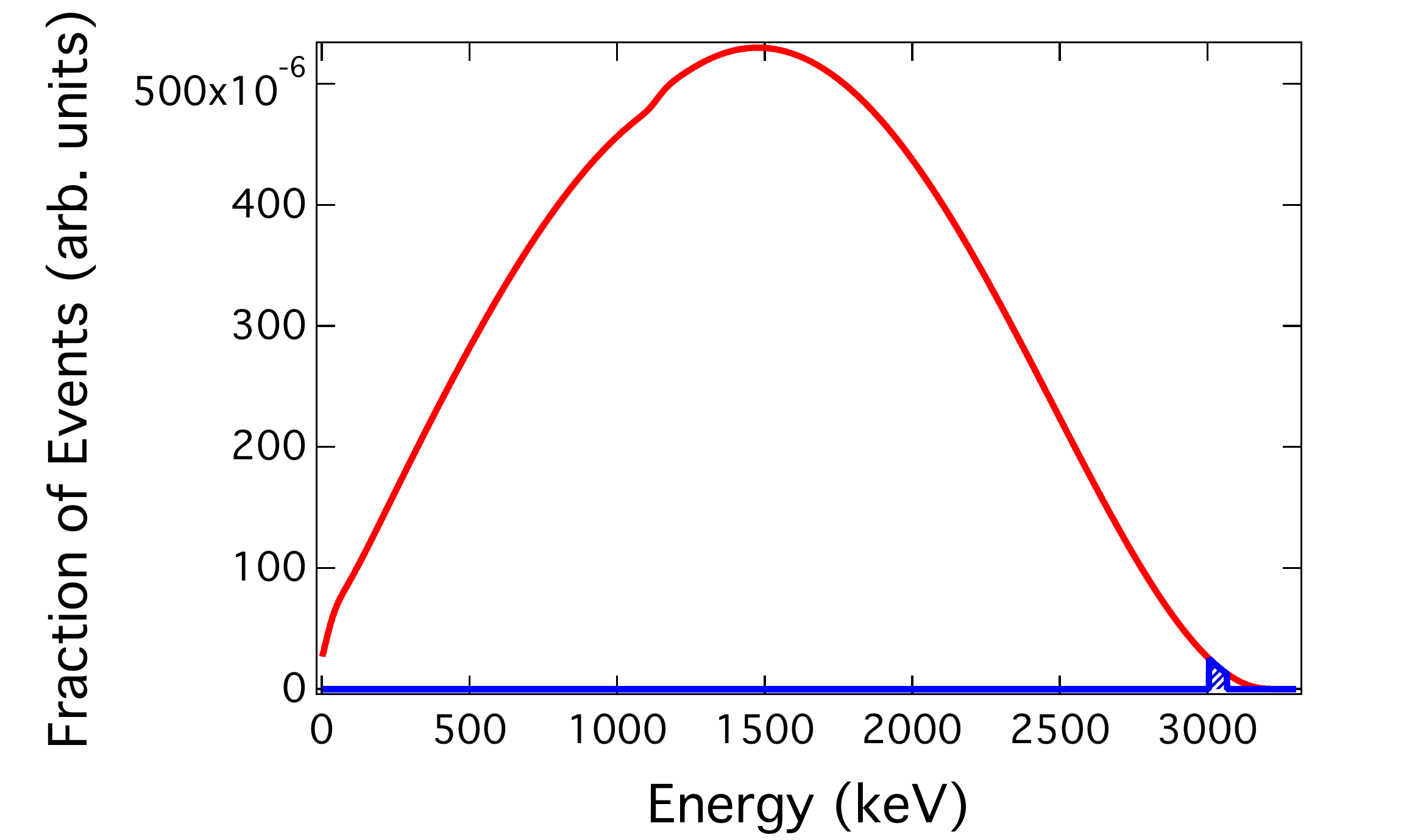}
\end{center}
\end{figure}

The oscillated solar $\nu$ capture rate on \nuc{100}{Mo} is calculated to be 539 SNU.  \nuc{100}{Tc} decays with a 93\% branching ratio of to the ground state of \nuc{100}{Ru} and 5.7\% to 1130-keV state. These were the only states considered in our calculations. The $\epsilon_{SB}$ for decay of the \nuc{100}{Tc} product to populate the ROI was calculated to be 0.06\%, 0.1\% and 0.3\% for resolutions of $\delta$ = 0.01, 0.02 and 0.05 respectively. The resulting $SB$ spectrum is shown in Fig.~\ref{fig:Tc100}.

\subsubsection{\nuc{150}{Nd}}
\begin{figure}[htb]
\caption{The sum spectrum of $\beta$ and $\gamma$-ray energies for the decay of \nuc{150}{Pm}. The hatched region shows the ROI fraction for a $\delta = 0.02$.
\label{fig:Pm150}. }
\begin{center}
\includegraphics[width=9 cm]{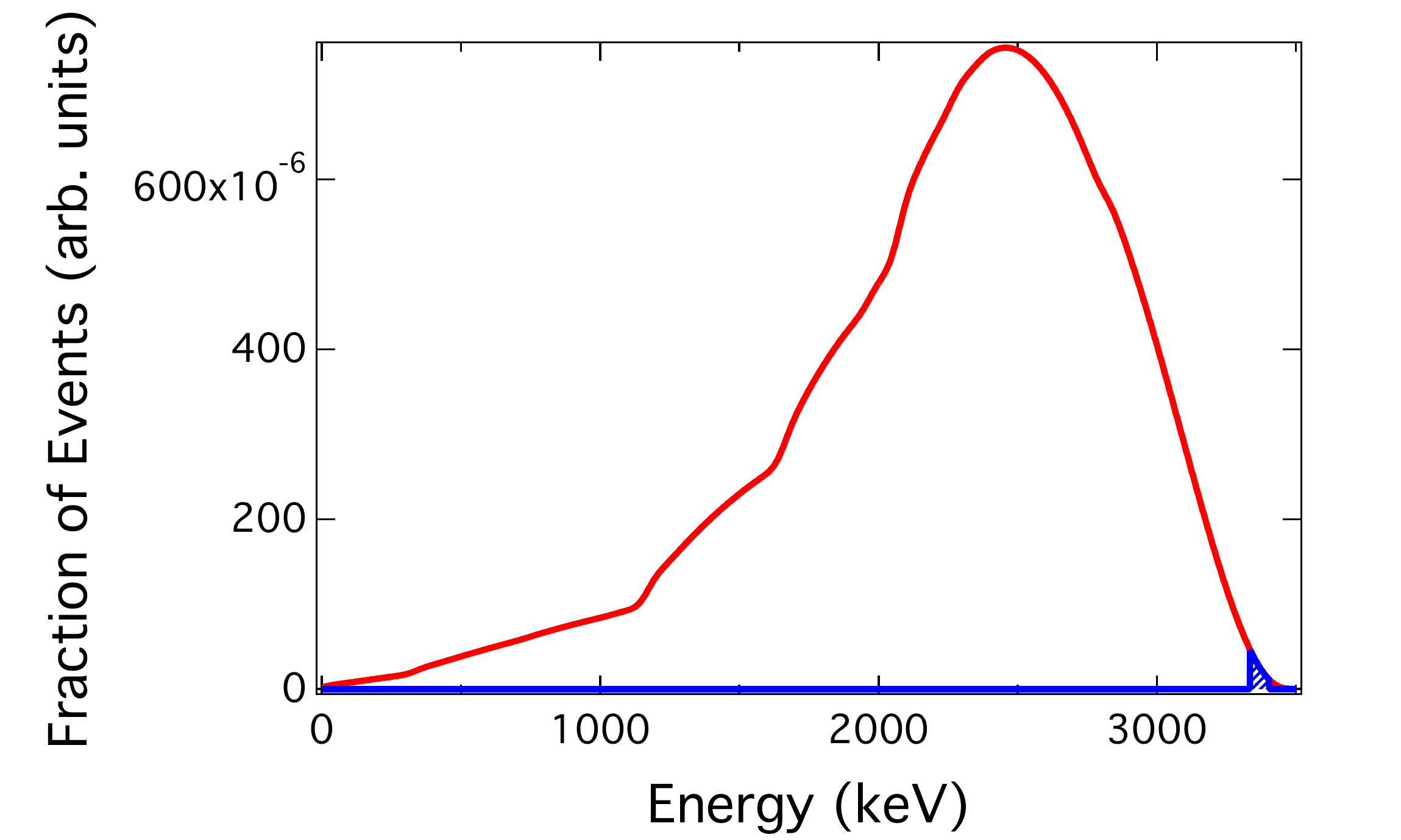}
\end{center}
\end{figure}

The oscillated solar $\nu$ capture rate on \nuc{150}{Nd} is calculated to be 524 SNU.  \nuc{150}{Pm} decays to numerous excited states in \nuc{150}{Sm}. We included all states to a branching ratio of 1\% for a total branching ratio of 99.7\%. However, the branching ratio to the ground state is uncertain and we took its value to be the upper limit value of 10\%. There were 12 states included in our calculations. The $\epsilon_{SB}$ for decay of the \nuc{150}{Pm} product to populate the ROI was calculated to be 0.08\%, 0.2\% and 0.7\% for resolutions of $\delta$ = 0.01, 0.02 and 0.05 respectively. The resulting $SB$ spectrum is shown in Fig.~\ref{fig:Pm150}.

\subsection{The Solar Neutrino CC Interaction}

The \nuc{8}{B} $\nu$ spectrum extends to energies higher than $Q_{\beta \beta}$. Therefore energy deposits from \nuc{8}{B} solar-${\nu}$ CC interactions at the ROI are possible. Hence, we consider CC reactions to the i$^{th}$ GT state in $^Z$A at energy $E_i$ above the ground state with e$^-$ emission followed by $\gamma$ decays to the ground state in $^Z$A. For simple calorimetric detectors, the sum energy of the CC produced e$^-$ energy ($E_e$) and any emitted $\gamma$-ray energy ($E_{\gamma }$) is measured. If the sum of these energies lies within the ROI at $Q_{\beta \beta}$, it will be a background to \BBz. The resulting values for $E_{\nu}$ and $E_e$ are obtained from the condition, $E_e + E_i=Q_{\beta \beta}$, as;
\begin{equation}
E_{\nu}= Q_{\beta \beta }+Q_{e},~~E_e=Q_{\beta \beta }-E_i,
\end{equation}
where $Q_e$  is  the electron capture Q value for the ground state of $^{Z}$A.  Note for \nuc{82}{Se}, $Q_e$ =  171.7 keV is the Q value for $\nu$ capture to the 75.1 keV meta-stable state in $^{82}$Br, not the ground state as for the other nuclei under consideration here and correspondingly for this case, $E_i$ is the energy from the meta stable state. The 5$^-$ spin-parity assignment of the $^{82}$Br ground state results in a negligible contribution to the CC interaction.
The background rate due to $^8$B-$\nu$ capture is obtained as 
\begin{equation}
B_{CC}=\phi (^8B,E_{\nu})\Delta E \sum_i\sigma(E_i),
\end{equation} 
where $\phi(^8B,E_{\nu})$ is the $^8$B-$\nu$ flux per MeV at $E_{\nu}$ and $\sigma(E_i)$ is the neutrino CC cross section \cite{eji14} to excite the i$^{th}$ GT state at $E_i$, and the sum extends over all GT states with $E_i \ge Q_{\beta \beta }$. $B_{CC}$ is  proportional to the fractional resolution $\delta =\Delta E/Q_{\beta \beta}$ because the fraction of the $^8$B-$\nu$ flux contributing to the ROI is proportional to the width of the energy window. 

These  $B_{CC}$ rates are smaller by 2-3 orders of magnitude than $B_{SB}$ rates as given in Table~\ref{tab:SolarRates}. Thus one may ignore the background contribution of $B_{CC}$ for the present nuclei of the Group A as was done for the solar-$\nu$ backgrounds for the nuclei in the Group B \cite{eji14}.

\section{Summary and Discussion}
The evaluated background rates for simple calorimetric detectors given in Table~\ref{tab:SolarRates} are approximately 220 $\delta$/t y for $^{82}$Se, 5.5 $\delta $/t y for $^{100}$Mo and 6.0 $\delta$/t y for $^{150}$Nd. In case of $\delta =0.02$, they are, respectively, 4.42/t y, 0.11 /t y and 0.12/t y. For the case of $\delta$=0.05, they are, respectively, 11/t y, 0.28/t y and 0.30/t y. Thus solar-$\nu$ backgrounds are serious for $^{82}$Se experiments proposing to study the IH $\nu$ mass region unless the fractional resolution is reduced below 10$^{-3}$. 

For experiments with modest resolution, \BBt\ may also be a background at the ROI. Since this background ($B_{2\nu}$) and that of $B_{SB}$ both depend on resolution, we compare the two. Using an approximation~\cite{ell02} for the number of \BBt\ events that populate in the ROI, the number of background events for 2\% fractional resolution is also given in Table~\ref{tab:SolarRates}. Because the \BBt\ half lives vary by a factor of $\approx$300 for these 6 isotopes, this background varies greatly also. For \nuc{100}{Mo} and \nuc{150}{Nd}, and to a lesser extent \nuc{82}{Se}, this is a significant issue for experiments approaching the ton scale and $B_{2\nu}$dominates over $B_{SB}$. Therefore, the resolution of $^{100}$Mo and $^{150}$Nd detectors should be around 0.01 or less to avoid serious background from \BBt\ to study the inverted hierarchy $\nu$ mass region. For \nuc{76}{Ge}, \nuc{130}{Te}, and \nuc{136}{Xe} $B_{2\nu}$ is small. 

It is noted that fractional resolution of around 2\% or less is required for the $^{130}$Te and $^{136}$Xe  experiments to reduce $B_{SB}$ in order to study the inverted hierarchy mass region.

So far, we have discussed selection of the true \BBz\ signal and rejection of the background by only energy selection at the ROI in simple calorimetric detectors. Improvement of the energy resolution reduces the background rates by reducing the width $\Delta E$ of the ROI energy window. There are also several techniques to reduce $\epsilon_{SB }$ and hence $B_{SB}$. Each technique depends on specific detector configurations and thus we do not discuss such details in the present paper, but briefly describe possible reductions in general.

For solar-$\nu$ interactions on $^{82}$Se and $^{150}$Nd, the intermediate nuclei of $^{82}$Br and $^{150}$Pr decay primarily by emitting $\beta$ and $\gamma$ rays. One can reduce $B_{SB}$ by measuring the spacial distribution of the energy deposits, since $\gamma$ rays will interact in a much larger volume of the detector than the \BBz\ signal. If detector segments are small, the $\gamma$ rays detected outside the segment can be used to reject $B_{SB}$ and $B_{CC}$.  

Signal selection by time correlation (SSTC) can be used to reduce solar-${\nu}$ background rates \cite{eji05}. Since the half-life of the intermediate nucleus of $^{100}$Tc is only 16 sec, $B_{SB}$ in $^{100}$Mo \BB\ experiments can be rejected by delayed anti-coincidence with the preceding CC e$^-$.  This technique, however, is not useful for $^{82}$Se and $^{150}$Nd because of the long half-lives (35.3 hr  and 2.68 hr, respectively) for the intermediate nuclei.  In other wards,  $^{100}$Mo can be used to study pp solar-$\nu$ CC e$^-$ by a delayed coincidence with the successive SB $\beta$ rays \cite{eji00}, but $^{82}$Se and $^{150}$Nd  would be less effective as solar $\nu$ experiments. 

Tracking detectors as used for real-time $^{82}$Se and $^{100}$Mo \BB\ experiments can be used to select \BB\ signals by measuring the individual two $\beta $ rays \cite{eji01,arn05}, and to reject $B_{SB}$ with only one $\beta$ ray. However, development of  large tracking chambers with multi-ton scale enriched \BB\ isotopes is a real challenge for future.    

Finally, it should be remarked that the $\nu$-mass sensitivity for \BB\ experiments with $NT$ t y of \BB\ isotope exposure is given by~\cite{eji05,ver12,ell02};
\begin{equation}
 m_{\nu} < \frac{7.8 \mbox{~meV}}{M_{0\nu}} \sqrt{\frac{A}{G\epsilon}} \left( \frac{B}{NT} \right)^{1/4}
\end{equation}
with constant $G$, $A$, $M_{0\nu}$, $\epsilon$, $B$ being the phase space in units of $10^{-14}$/y, mass number, nuclear matrix element, detector efficiency, and background rate per year in the ROI per ton of \BB\ nuclei.  Therefore one needs to optimize a number of key elements when planning future \BB\ experiments with the background being critical. The solar-$\nu$ background discussed here is just one of many components of the background, that needs to be considered for high-sensitivity \BBz\ experiments hoping to cover the IH $\nu$-mass region.\\

\section*{Acknowledgments}
We acknowledge support from the Office of Nuclear Physics in the Department of Energy, Office of Science. We gratefully acknowledge the support of the U.S. Department 
of Energy through the LANL/LDRD Program.


\begin{thebibliography}{9} 
      
\bibitem{ell04}
S. Elliott S and J. Engel  2004 {\it J. Phys. G. Nucl. Part. Phys.} {\bf 30} R183  
\bibitem {eji05} 
H. Ejiri 2005 {\it J. Phys. Soc. Japan} {\bf 74} 2101
\bibitem{avi08}
F.T. Avignone III, S.R. Elliott S and J. Engel  2008 {\it Rev.Mod. Phys.} {\bf 80} 481   
\bibitem {ver12} 
J. Vergados, H. Ejiri, F. {\v S}imkovic 2012 {\it Rep. Prog. Phys. } {\bf 75}  106301
\bibitem{eji14} 
 H. Ejiri and S. Elliott 2014 {\it Phys. Rev. C}  {\bf 89} 055501
 \bibitem{bar11}
N.F. de Barros and K. Zuber  2011 {\it J. Phys}. G {\bf 38} 105201
 \bibitem{eji00}
H. Ejiri et al. 2000 {\it Phys. Rev. Lett.} {\bf 85} 2917. 
\bibitem{zub03}
K. Zuber  2003 {\it Phys. Lett. B} {\bf 571} 148
\bibitem{zub12}
K. Zuber  2012 {\it Phys. Lett. B} {\bf 709}  6
\bibitem{bar14}
N.F. de Barros, J. Thurn, K. Zuber  2014 {\it J. Phys}. G {\bf 41} 115105
\bibitem{eji16}
H. Ejiri and K. Zuber 2016 {\it J. Phys. G} {\bf 43} 045201.
\bibitem{eji78}
H. Ejiri and J.I. Fujita 1978 {\it Phys. Rep.} {\bf 38}C 85. 
\bibitem{fre16}
D. Frekers, et al. 2016 {\it Phys. Rev. C} {\bf 94} 014614.
\bibitem{thi12}
J.H. Thies et al., 2012 {\it  Phys. Rev. C} {\bf 86} 044309.
\bibitem{gue11}
C.S. Guess et al., 2011 {\it Phys. Rev. C} {\bf 83} 064318.
\bibitem{bah05} 
J.N. Bahcall, A.M. Serenelli, and S. Basu, Astrophys. J. Lett. 621, L85 (2005).

\bibitem{bar15}
A.S. Barabash, Nucl. Phys. A {\bf 935} 52, (2015).
\bibitem{ago15}
M. Agostino et al., Eur. Phys. J. C {\bf 75} (2015) 416.

\bibitem{eji01}
H. Ejiri et al. 2001 {\it Phys. Rev.} C {\bf 63} 065501.
\bibitem{arn05}
C. Arnaboldi et al. 2005 {\it Phys. Rev. Lett.} {\bf 95} 142501. 
\bibitem{ell02}
S.R. Elliott and P. Vogel, Annu. Rev. Nucl. Part. Sci. {\bf 52}, 115 (2002).

\end{thebibliography}
\end{document}